\documentclass[%
 reprint,
 amsmath,amssymb,
 aps,
 prl,
]{revtex4-2}

\usepackage{graphicx}
\usepackage{dcolumn}
\usepackage{bm}
\usepackage{xcolor,colortbl}
\usepackage{multirow}
\usepackage{longtable}
\usepackage{array}
\usepackage{lineno}

\definecolor{Grey}{gray}{0.9}
\definecolor{LightRed}{rgb}{0.937,0.675,0.639}
\newcolumntype{x}[1]{>{\centering\let\newline\\\arraybackslash\hspace{0pt}}p{#1}}

\begin{document}

\preprint{APS/123-QED}

\title{Rising and Sinking in Resonance: 
Probing the critical role of rotational dynamics for buoyancy driven spheres}

\author{Jelle B. Will}
\email{j.b.will@utwente.nl}
\author{Dominik Krug}%
\email{d.j.krug@utwente.nl}
\affiliation{Physics of Fluid Group and Max Planck Center Twente, J. M. Burgers Centre for Fluid Dynamics, University of Twente, P.O. Box 217, 7500 AE Enschede, The Netherlands.
}

\date{\today}

\begin{abstract}
We present experimental results for spherical particles rising and settling in a still fluid. Imposing a well-controlled center of mass offset enables us to vary the rotational dynamics selectively by introducing an intrinsic rotational timescale to the problem. Results are highly sensitive even to small degrees of offset, rendering this a practically relevant parameter by itself. We further find that for a certain ratio of the rotational to a vortex shedding timescale (capturing a Froude-type similarity) a resonance phenomenon sets in. Even though this is a rotational effect in origin, it also strongly affects translational oscillation frequency and amplitude, and most importantly the drag coefficient. This observation equally applies to both heavy and light spheres, albeit with slightly different characteristics for which we offer an explanation. Our findings highlight the need to consider rotational parameters when trying to understand and classify path properties of rising and settling spheres.
\end{abstract}

\maketitle

A particle rising or settling in a fluid otherwise at rest is a fundamental problem in fluid mechanics. Despite its apparent simplicity, the problem is remarkably complex as the coupling between the motion of the body and the surrounding flow field results in an exceedingly wide range of complex trajectories \cite{Lamb:1932,Lugt1983,bearman1984,williamson2004vortex,Ern:2012,mathai2020}. The dynamics are dependent on the geometry and mass of the body as well as the fluid properties. In particle-laden flows, single particle dynamics often persist \cite{Magnaudet2000} and can significantly affect global properties of a system such as sedimentation rate, and transport of heat or nutrients in a fluid \cite{zhao2010}, or mixing for chemical reactors \cite{almeras2015,almeras2019}. Therefore, a fundamental understanding of the behaviour of individual particles is of primary importance in understanding larger systems in nature and industrial applications.

However, even for the most fundamental geometry, i.e. a sphere, the observed behaviour is not completely explained and understood \cite{Auguste2018,Mathai2018}. The traditional notion is that the two-way coupled dynamics for this case depend on two dimensionless parameters only: the particle-to-fluid mass density ratio $\Gamma \equiv \rho_p/\rho_f$, and the particle Galileo number $\textrm{Ga} \equiv U_b D/\nu$ \cite{Jenny2003,horowitz2008}. Here, $D$ is the particle diameter, $\nu$ the kinematic viscosity of the fluid, and $U_b = \sqrt{|1-\Gamma|g D}$ is the buoyancy velocity with $g$ denoting the acceleration due to gravity. Out of these parameters, $\Gamma$ accounts for the particle response to pressure fluctuations in the flow (inertial effects) and, by relating buoyancy and viscous forces, Ga is indicative of the flow structure around the body \cite{Jenny2003}. A closely related quantity to the latter is the 
Reynolds number $Re \equiv \langle u_z \rangle D/ \nu$, where $\langle u_z \rangle$ is the mean vertical velocity (with $\langle \cdot \rangle$ denoting a time and ensemble average) which is not known \emph{a priori,} however.

A significant amount of work was aimed at classifying the motion of spheres and differences in their wake structures as a function of $\Gamma$ and Ga. \cite{Jenny2004,veldhuis2004,veldhuis2007,horowitz2008,Horowitz:2010,Auguste2018}. However, there still exists substantial disagreement even on fundamental aspects. For example, it remains open why there are conflicting results for the parameter range for which strong path oscillations are observed \cite{preukschat1962,shafrir1965,christiansen1965,Karamanev:1992,karamanev1996,Jenny2003,Jenny2004,veldhuis2004,veldhuis2007,veldhuis2009,Horowitz:2010}. The lack of a universal description alludes to the possibility that additional -- yet largely unexplored -- parameters may play a role in the problem. In fact, recently, the importance of rotational dynamics for spheres and 2D cylinders has been highlighted \cite{namkoong2008,Mathai:2017,Mathai2018}, showing that the moment of inertia (MoI) can affect the vortex shedding mode, the frequency and amplitude of oscillation, and the vertical velocity. The key physical mechanism behind this rotational-translational coupling is the Magnus lift force, which in a still fluid is given by $\boldsymbol{F}_{m}\sim \boldsymbol{\omega} \times \boldsymbol{u}$  \cite{loth2008lift}, with  $\boldsymbol{\omega}$ and $\boldsymbol{u}$ denoting the particle angular and linear velocity vectors, respectively. It has been suggested that the dependence on the particle MoI can be one of the factors contributing to the spread in particle drag coefficient as well as causing differences in oscillation amplitude \cite{Mathai2018}, but conclusive evidence, in particular for spheres, is missing.

\begin{figure}[b]
\includegraphics[width=0.45\textwidth]{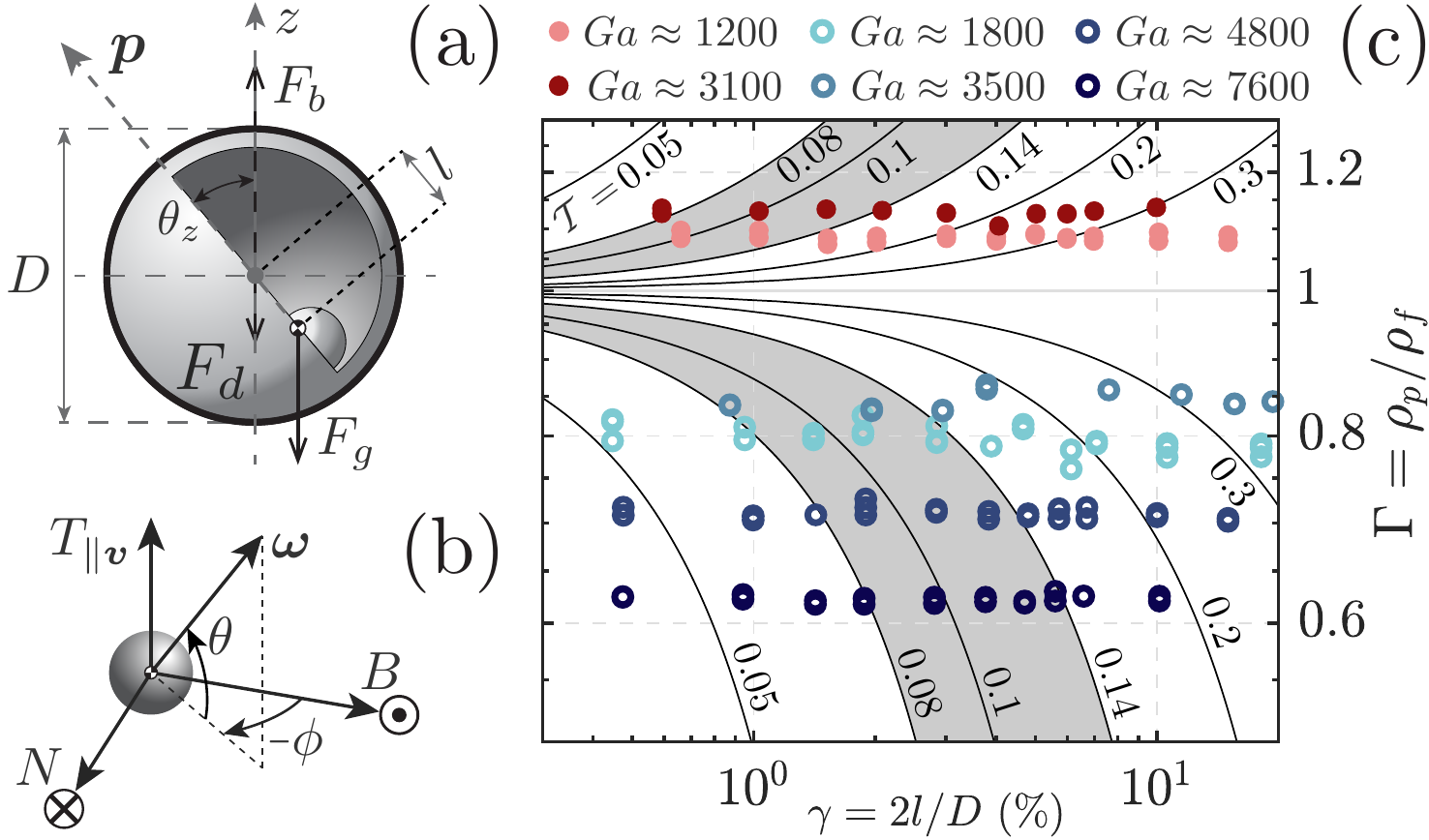}
\caption{\label{fig:particles} (a) Schematic of a sphere with CoM offset; the metal sphere inside the shell is displaced in the radial direction along $\boldsymbol{p}$, the unit vector pointing from the CoM (checkered circle) to the geometrical centre (grey circle). The distance between these two points is the offset $l$. (b) The particle Frenet–Serret (TNB) coordinate system, with unit vectors $T$ (parallel to $\boldsymbol{u}$), $\boldsymbol{N}$ (pointing in the direction of curvature of the path), and $\boldsymbol{B}$ (defined such that $\boldsymbol{N} = \boldsymbol{B} \times \boldsymbol{T}$).
The angles $\phi$ (azimuth) and $\theta$ (elevation) uniquely define a vector in this space. (c) Explored parameter space. Grey shading indicates the resonance regime and $\mathcal{T}$-isocontours correspond to $I^* = 1$.}
\end{figure}

In this Letter, we systematically explore the effect of rotational dynamics on rising and settling spheres. To this end, we modify the rotational properties of the spherical particles in a controlled manner by introducing a center of mass (CoM) offset $\gamma \equiv 2l/D$, where $l$ is the distance along the unit vector $\boldsymbol{p}$ pointing from the geometrical centre to the CoM (see Fig.\,\ref{fig:particles}(a)). Clearly, such an offset can also be expected to occur in a host of practical applications, where particle properties are rarely ever uniform. This concerns e.g. the falling of dandelion seeds \cite{cummins2018} and snowflakes \cite{nemes2017,li2020,zeugin2020,mccorquodaletrail_p1,mccorquodaletrail_p2}, the sedimentation behaviour of sand grains and stones \cite{richardson1954,meiburg2010}, chemical and biological reactors with (inverse) fluidized beds \cite{sowmeyan2008}, as well as the transport of micro-plastic in the oceans \cite{clark2020settling}. The practical relevance is moreover rooted in the fact that we find that even small values of $\gamma$ can affect the kinematics and dynamics of spherical particles significantly. Despite their apparent relevance, CoM offsets are often listed more generally as potential sources of experimental uncertainty (e.g. \cite{Ern:2012}) but only few studies have considered $\gamma$ explicitly.
To our knowledge, the relevance of this parameter was first noted by \cite{Jenny2004} who report that the trajectory of a settling sphere with Ga $= 180$ was destabilized when introducing an offset of $\gamma = 5\%$ (originating from an air bubble trapped in some of their particles). More recently, it was shown that lateral motion of spheres in a linear shear flow was reduced by presence of a strong offset \cite{tanaka2020}. While both of these studies clearly underline the relevance of $\gamma$ as a parameter, the accounts remain anecdotal and a complete picture based on a systematic variation is lacking still. 
For completeness, it should be mentioned that the role of mass asymmetry has also been examined in the context of cylindrical or fibre like particles \cite{yasseri2014,angle2019,roy2019}. However, due to the anisotropic geometry, the dynamics in these instances are completely different to the spherical case considered here.

We start our analysis from the classical Kelvin-Kirchhoff equations \cite{mougin2002}. These express the conservation of linear and angular momentum in a reference frame  rotating with the particle but with a fixed origin. For a suspended sphere, they are given by:
\begin{equation}
\left( 1 +\dfrac{1}{2\Gamma}  \right) \left( \dfrac{\textrm{d} \boldsymbol{u}}{\textrm{d} t} + \boldsymbol{\omega} \times \boldsymbol{u} \right) = \dfrac{\boldsymbol{F}_f}{m_p} + \dfrac{(1-\Gamma) g }{\Gamma}\boldsymbol{e}_z, \label{eq:KKF}
\end{equation}
\begin{equation}
\dfrac{1}{10}I^* \dfrac{\textrm{d} \boldsymbol{\omega}}{\textrm{d} t} = \dfrac{\boldsymbol{T}_f}{m_p D^2} - \dfrac{\gamma}{2D}   (g\boldsymbol{e}_z + \boldsymbol{a}_c) \times \boldsymbol{p}.\label{eq:KKT}
\end{equation}
 Here, $\boldsymbol{F}_f$ and $\boldsymbol{T}_f$ are the fluid force and torque applied to the body, respectively, and $\boldsymbol{e}_z$ is the vertical unit vector. Further, we define the dimensionless MoI $I^* \equiv I_p/I_\Gamma$ as the ratio of the particle MoI over the MoI of a sphere with a uniform density distribution $I_\Gamma = 1/10m_p D^2$, where $m_p$ is the particle mass. Note that the linear momentum balance (Eq.\,\ref{eq:KKF}) remains unaffected by the choice of $\gamma$.  Eq.\,\ref{eq:KKT} represents the angular momentum balance around the center of the sphere, in which the effect of the CoM offset appears in the form of the cross-product on the right-hand side. Apart from $\gamma$, the magnitude of this term also depends on the included angle $\theta_z$ between $\boldsymbol{p}$ and $\boldsymbol{e}_z$ (see Fig.\,\ref{fig:particles}(a)), and on $\boldsymbol{a}_c$, the acceleration of the center of mass.

For spheres, the geometric center and the center of pressure coincide. Therefore, the forcing term $\boldsymbol{T}_f$ in Eq.\,\ref{eq:KKT} is solely due to skin friction, which at high Re is approximately periodic and associated with the vortex shedding in the wake of the body \cite{govardhan2004}. Neglecting the additional dependence on $\boldsymbol{a}_c$, the offset term acts as a restoring torque. Thus, Eq.\,\ref{eq:KKT} is similar to a periodically forced pendulum with a natural frequency $f_p = \sqrt{5 \gamma g/D I^*}/2 \pi$, and the corresponding timescale $\tau_p = f_p^{-1}$. The driving due to vortex shedding is characterised by $\tau_v \sim D/U_b$ and on this basis, we define the ratio
\begin{equation}
\mathcal{T} = \dfrac{\tau_v}{\tau_p} = \dfrac{1}{2\pi} \sqrt{\dfrac{5\gamma}{|1-\Gamma|I^*}} \label{eq:Fr}.
\end{equation}
Note that $\mathcal{T}$ is entirely determined by particle properties. In relating translational ($U_b$) and dissipative ($D/\tau_p$) velocities, $\mathcal{T}$ corresponds to the inverse of the Froude number defined in \cite{Belmonte:1998} for falling strips. However, the definition in Eq.\,\ref{eq:Fr} is preferred here as it avoids divergence at $\gamma = 0$. 

\begin{figure*}
\includegraphics[width=1\textwidth]{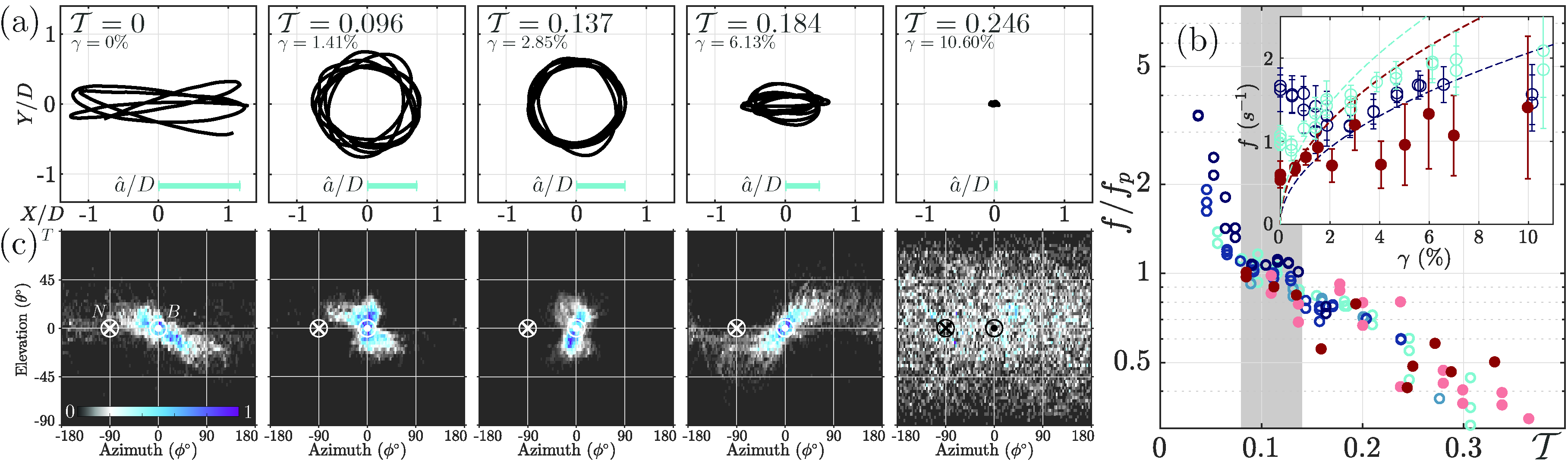}
\caption{\label{fig:traj} (a) Characteristic particle trajectories as seen from the top for different values of $\mathcal{T}$ at $\textrm{Ga} \approx 1800$, the length of the horizontal blue lines represent the corresponding amplitudes $\hat{a}/D$. (b) The inset shows the frequency of the horizontal path oscillation as a function of $\gamma$ for 3 different Ga values. Dashed lines of corresponding color represent the respective pendulum frequency $f_p(\gamma)$. The main figure shows the ratio  $f/f_p$  vs. $\mathcal{T}$ for the entire dataset. (c) Normalized histograms of the orientation of  $\boldsymbol{\omega}$ in the TNB coordinates introduced in Fig\,\ref{fig:particles}\,({\it b\/}). The instantaneous direction of motion ($\boldsymbol{T}$) here corresponds to $\theta = 90^\circ$. The cross marker at $\phi = 0^\circ$ $\theta = 0$ indicates the direction of acceleration ($\boldsymbol{N}$), the dotted circle at $\phi = \theta = 0$ indicates the $\boldsymbol{B}$-direction.}
\end{figure*}

To test the effect of variations in $\mathcal{T}$, laboratory experiments were performed for rising and settling spheres in a still fluid with systematic variations in Ga, $\gamma$, and $\Gamma$. 
At a given Ga and $\Gamma$, the variation $I^*(\gamma)$ was below $ 2.5\%$ of $I^*(\gamma = 0)$ for moderate offsets ($\gamma < 5\%$) and only exceeded 10\%  of that value in the most extreme cases. An overview over the explored range of parameters is shown in Fig.\,\ref{fig:particles}(c), where isocontours of $\mathcal{T}$ are included for $I^* = 1$. Notably, our experiments include both heavy ($\Gamma >1$) and light ($\Gamma <1$) spheres and in all cases $\Gamma > 0.61$, the upper limit for the existence of path oscillations according to \cite{Horowitz:2010}.

Particles were left to settle or rise in a large vertical water tank. After an initial transient, the position and orientation of the spheres were tracked with optical methods \cite{Mathai:2016,will2020kinematics}. Details of the setup and the postprocessing of the data are provided in the supplementary materials, that also include movies of rendered trajectories at $Ga =1800$.

The profound effect variations in $\gamma$ have on particle kinematics is exemplified in Fig.\,\ref{fig:traj}(a), where horizontal projections ($XY$-plane) of drift corrected trajectories for the $\textrm{Ga} \approx 1800$ (rising) case are shown. From these plots, it is obvious that the oscillation amplitude varies significantly with $\gamma$ and even vanishes for the most extreme offset. Simultaneously, also the shape of the oscillations transitions from  mostly planar to circular and then back to a more planar motion with additional precession as $\gamma $ is increased. A similar behaviour is observed across all Ga and $\Gamma$ for rising particles. The increase in amplitude also occurred for $\Gamma >1$, but not the more helical trajectories or precession at higher $\mathcal{T}$. We did not encounter significant horizontal drift, as is reported for lower Ga \cite{Auguste2018}, for any of the cases considered here.

As a first quantitative measure, we extract the frequency $f$ of the horizontal path oscillations. Sample results for three cases in the inset of Fig.\,\ref{fig:traj}(b) reveal that $f$ varies significantly with $\gamma$ with a remarkable sensitivity even at small offsets. All cases display a similar pattern relative to their respective pendulum frequency $f_p(\gamma)$ (dashed lines): At small $\gamma$, $f$ exceeds $f_p$ but the two quickly converge as the offset is increased resulting in a resonance ($f \approx f_p$) between the path oscillations (and hence the vortex shedding) and the rotational dynamics of the particle. For offsets greater than those at resonance, $f_p$ quickly outgrows the shedding frequency and path oscillations damp out (resulting in the large error bars in the experimental data).
Resonance occurs at different values of $\gamma$ for different particles. However, all data collapse when plotting $f/f_p$ against $\mathcal{T}$ as is done in the main panel Fig.\,\ref{fig:traj}(b). This confirms that $\mathcal{T}$ is indeed the relevant parameter governing the behaviour of particles with CoM offset and we identify the resonance range as $0.08 \lessapprox \mathcal{T} \lessapprox 0.14$ (marked by a grey shading in all figures). It is worth pointing out that a similar lock-in phenomenon of the wake to object oscillations was earlier observed for forced  translational oscillations of cylindrical or square beams in a cross flow \cite{bishop1964lift,bearman1982experimental}. A key difference and a remarkable feature of the present results is, however, that here vortex shedding dynamics are governed by a parameter that is intrinsically rotational.

\begin{figure*}
\includegraphics[width=1\textwidth]{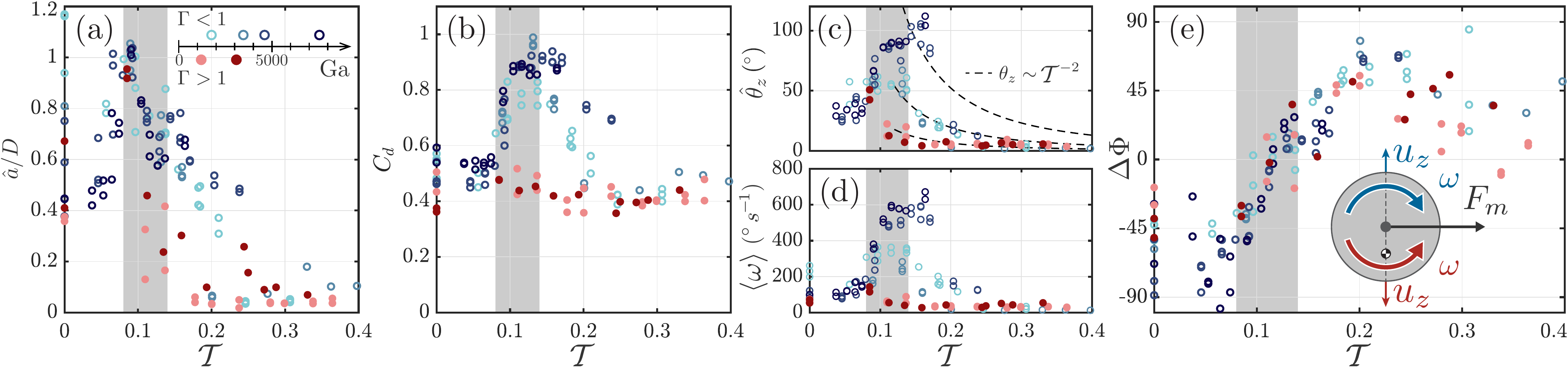}
\caption{\label{fig:Fr_compiled} Dependence on $\mathcal{T}$ for (a) amplitude of the path oscillations $\hat{a}/D$ (b) particle vertical drag coefficient $C_d$ (c) particle rotational amplitude $\hat{\theta}_z$ (d) time averaged angular velocity $\langle \omega \rangle$  (e) phase angle $\Delta \Phi$ between the horizontal particle acceleration and the Magnus lift force. For all quantities, datapoints represent averages over multiple experiments with the same particle.}
\end{figure*}

The resonance behaviour revealed for the frequencies also has a direct imprint on other parameters, such as the normalized oscillation amplitude $\hat{a}/D$ shown in Fig.\,\ref{fig:Fr_compiled}(a) for both heavy and light particles. At $\mathcal{T} = 0$, the scatter in $\hat{a}/D$ is considerable owing to the variation in Ga, $I^*$ and  $\Gamma$. However, these differences vanish and the variation of $\hat{a}/D$ as function of $\mathcal{T}$ becomes remarkably similar across all cases tested rendering this the dominant parameter once a small but finite offset ($\gamma>0$) is introduced. Amplitudes are largest in the resonance band with a peak of $\hat{a}/D \approx 1$ located at $\mathcal{T}\approx 0.09$ for both rising and settling particles. Consistent with the observation in Fig.\,\ref{fig:traj}(a), path oscillations vanish at large $\mathcal{T}$ in all cases and it appears that the decrease in $\hat{a}/D$ beyond resonance is steeper for larger values of $\Gamma$. While the resonant behaviour in terms of $f/f_p$ and $\hat{a}/D$ is very similar for heavy and light particles, remarkably the same is not true for the drag coefficient $C_d = 4 D | 1 - \Gamma|g/3\langle v_z \rangle_t^2$ shown in Fig.\,\ref{fig:Fr_compiled}(b). For rising spheres, there is almost a factor of two increase in $C_d$ in the resonance regime as compared to the $\mathcal{T}= 0$ case. In contrast, the $C_d$ results appear virtually insensitive to any changes in $\mathcal{T}$ for settling spheres.

A clue pointing to the cause of this surprising behaviour is given by the results for the rotational amplitude $\hat{\theta}_{z}$ in Fig.\,\ref{fig:Fr_compiled}(c). The resonance peak for $\hat{\theta}_{z}$ is prominent at low $\Gamma$ reaching values even beyond 90$^\circ$, but remains weak for $\Gamma>1$. In all cases, the rotational amplitude vanishes largely for higher $\mathcal{T}$, for which $f<f_p$. Indeed, the scaling $\hat{\theta}_z \sim \mathcal{T}^{-2}$, which follows from a quasi-static assumption using $T_f\sim \rho_f D^3 U^2$ \cite{jordan1972,bouchet2006} appears to capture the decay of $\hat{\theta}_z$ with increasing $\mathcal{T}$ well in this regime. Such a simple argument fails, however, to reproduce the prefactor properly for which the suggested $(\Gamma I^*)^{-1}$-dependence is weaker than the actual variation in the data.
Dynamically, the rotation rate is more relevant than the rotational amplitude due to its relation to the Magnus lift force. It further provides a more robust measure, even at zero offset. We therefore additionally consider the mean rotation rate $\langle \omega \rangle$ in Fig.\,\ref{fig:Fr_compiled}(d) and observe a good agreement between the trend of this quantity and that of $C_d$ as a function of $\mathcal{T}$. This indicates that instead of the path oscillation amplitude (which features a resonance peak even for $\Gamma >1$), the drag experienced by a rising or settling sphere correlates better with the amount of particle rotational kinetic energy that governs the drag experienced by a rising or settling sphere. 

In evaluating the nature of the rotational-translational coupling, it is useful to consider the Lagrangian Frenet-Serret coordinate system ($\boldsymbol{T}$,$\boldsymbol{N}$,$\boldsymbol{B}$, see Fig.\,\ref{fig:particles}(b)), which is defined with respect to the path of the sphere \cite{loth2008lift,zimmermann2011rotational,Mathai2018}.
In Fig.\,\ref{fig:traj}(c), we have included histograms of the orientation of $\boldsymbol{\omega}$ in the TNB coordinate frame corresponding to the sample trajectories displayed in  Fig.\,\ref{fig:traj}(a). Especially for the resonance cases ($\mathcal{T}= 0.096$ and $\mathcal{T}= 0.137$), $\boldsymbol{\omega}$ is found to align strongly with $\boldsymbol{B}$. This implies that the normal acceleration (along $\boldsymbol{N}$) is consistent with the direction of the Magnus lift force in this state, since $\boldsymbol{F}_m \sim \boldsymbol{\omega} \times \boldsymbol{u}$.
In addition to the fact that no significant path oscillations are observed in the absence of particle rotation at high $\mathcal{T}$ (Fig.\,\ref{fig:Fr_compiled}), this underlines the crucial role rotational dynamics play for the path oscillations. The alignment between $\boldsymbol{\omega}$ and $\boldsymbol{B}$ in the resonance range is generally a robust feature for all cases considered here. Deviations between the two orientations are somewhat larger for settling particles resulting in broader peaks of the histograms corresponding to resonance (see supplementary material), but our conclusions remain valid also for these cases. While light particles at $\mathcal{T}$ outside resonance display distinct alignments away from  $\boldsymbol{B}$, this is not observed at $\Gamma >1$ as rotational amplitude quickly vanishes in those cases (leaving random alignment). 

With the relevance of the driving via the Magnus force established, it is then possible to analyse the phase relation between a forcing parameter and a system response. We do so by evaluating the phase angle $\Delta \Phi$ between the projections of the acceleration $\boldsymbol{a}$ and of the Magnus lift force $\boldsymbol{F}_m$ along an arbitrary horizontal direction. By definition, the particle acceleration lags behind the Magnus lift forcing for $\Delta \Phi < 0$ and vice versa for $\Delta \Phi > 0$. The results for $\Delta \Phi$ in  Fig.\,\ref{fig:Fr_compiled}(e) display a collapse as a function of $\mathcal{T}$ with a zero-crossing (at $\mathcal{T} \approx 0.12\pm 0.01$) within the resonance band. The latter is in line with the findings in figure \ref{fig:traj}(c) and implies an enhancement of path oscillations through $\boldsymbol{F}_m$. A key feature of the resonance is therefore that rotational-translational coupling is coherent with other forcing (e.g. through pressure forces induced by vortex shedding), while the two are less correlated otherwise. Interestingly, $ \Delta \Phi \approx 0^\circ$ occurs at $\mathcal{T}\approx 0.12$, at which rotations are strongest, whereas the phase lag is non-zero at the peak in $\hat{a}/D$ ($ \Delta \Phi \approx -45^\circ$ at $\mathcal{T}\approx 0.09$).

At this point, the question remains, why settling particles with CoM offset have such pronounced deficit in rotational dynamics as compared to rising ones. 
An explanation for this is related to the difference in alignment  between the direction of offset $\boldsymbol{p}$ (always pointing up) and the mean direction of motion, that switches between rising and settling particles. A Magnus lift force in the same direction is therefore associated with rotations in opposite directions between the two cases, as the inset in Fig.\,\ref{fig:Fr_compiled}(e) shows. This is relevant, because the torque induced by the lateral acceleration due to $\boldsymbol{F}_m$ (proportional to $\gamma \boldsymbol{a}_c \times \boldsymbol{p}$, see Eq.\,\ref{eq:KKT}) then either enhances (rising particles) or counteracts (settling) the rotation rate $\boldsymbol{\omega}$. Rotational amplitudes are therefore suppressed for heavy particles via this mechanism. 
In the resonance regime $\boldsymbol{F}_m$ strongly aligns with the direction of normal acceleration $\boldsymbol{N}$, such that also translational accelerations due to other forces amplify the effect in this case.

\begin{figure}[b]
\includegraphics[width=0.45\textwidth]{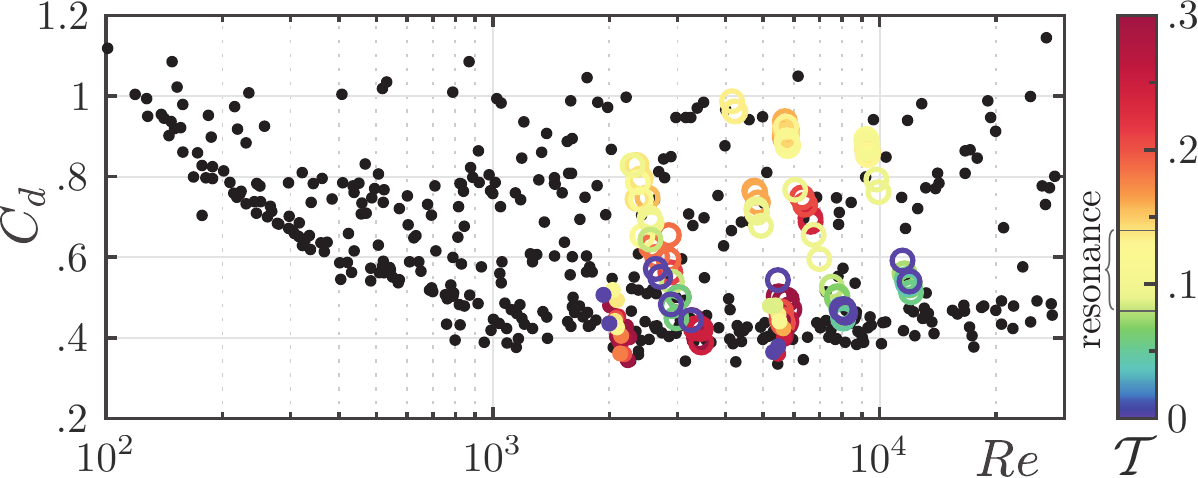}
\caption{\label{fig:cd_Re} Overview of particle drag coefficients of rising and settling spheres from literature (black dots) \cite{preukschat1962,maccready1964,shafrir1965,kuwabara1983,Jenny2004,karamanev1996,stringham1969,veldhuis2007,veldhuis2009,allen1900,liebster1927,lunnon1928,boillat1981,Horowitz:2010}, and results from the current data set, colour coded by $\mathcal{T}$, as a function of $\textrm{Re} = \langle u_z \rangle D/\nu$.}
\end{figure}

Finally, to put our results into perspective, we plot them alongside compiled literature data for spheres \cite{Horowitz:2010} in terms of $C_d$ versus $Re$ in Fig.\,\ref{fig:cd_Re}. The range of $C_d$ in the present measurements is seen to cover the full spread in the literature data. 
The fact that this variation arises from altering only the rotational dynamics is testament to the crucial importance of related parameters such as $I^*$ and $\gamma$. A complete description of the problem will therefore likely have to incorporate these. Our results further match the bounds of the literature data for $C_d$ very well, indicating that at least at this level the dynamics explored here are comparable to those encountered (nominally) without CoM offset. Moreover, there is a longstanding notion \cite{Horowitz:2010}, with mention already by Newton \cite{newton1999}, that high levels of $C_d$ are associated with large path amplitudes $\hat{a}/D$. This is clearly at odds with our results at $\Gamma > 1$ (but also with findings by others \cite{Jenny2004,zhou2015chaotic,Auguste2018,will2020kinematics}), where $C_d$ remains low even though $\hat{a}/D$ is significant. Our analysis suggests that $C_d$ is instead more closely related to particle rotations.

In summary, we have provided strong evidence for how critically the overall behaviour of free rising or sinking spheres is related to their rotational dynamics. The revealed sensitivity to CoM offsets as small as $\gamma = 0.5\%$ is remarkable and this parameter is therefore likely to play a role in many practical cases. In particular, it might affect the behaviour of spherical bubbles, which are known to display spiral or zigzag motion when rising in a contaminated liquid  \cite{haberman1953,saffman1956,hartunian1957}. In that case, a CoM offset might arise due to the fact that surfactants are swept to the back of the bubble by the flow and we estimate (assuming $\Gamma \to 0$ and $I^* = 1$) that $\gamma \approx 5\%$ would suffice to reach a $\mathcal{T}$-value in the resonance regime.
Clearly, the present findings are also useful to tailor particle behaviour.  In the future, it will be of particular interest to broaden the investigation to turbulent flow. Given how easily and effectively their resonance behaviour can be tuned, CoM spheres may be efficient means to `shape' turbulence by very selectively enhancing specific frequencies in the flow.\\

We thank Varghese Mathai, Chong Shen Ng, Chao Sun and Detlef Lohse for insightful discussions as well as Jim Scheefhals for assisting in the experiments. This work was supported by the Netherlands Organisation for Scientific Research (NWO)under VIDI Grant No. 13477.

\bibliography{References_complete}

\providecommand{\noopsort}[1]{}\providecommand{\singleletter}[1]{#1}%
\begin{thebibliography}{63}%
\makeatletter
\providecommand \@ifxundefined [1]{%
 \@ifx{#1\undefined}
}%
\providecommand \@ifnum [1]{%
 \ifnum #1\expandafter \@firstoftwo
 \else \expandafter \@secondoftwo
 \fi
}%
\providecommand \@ifx [1]{%
 \ifx #1\expandafter \@firstoftwo
 \else \expandafter \@secondoftwo
 \fi
}%
\providecommand \natexlab [1]{#1}%
\providecommand \enquote  [1]{``#1''}%
\providecommand \bibnamefont  [1]{#1}%
\providecommand \bibfnamefont [1]{#1}%
\providecommand \citenamefont [1]{#1}%
\providecommand \href@noop [0]{\@secondoftwo}%
\providecommand \href [0]{\begingroup \@sanitize@url \@href}%
\providecommand \@href[1]{\@@startlink{#1}\@@href}%
\providecommand \@@href[1]{\endgroup#1\@@endlink}%
\providecommand \@sanitize@url [0]{\catcode `\\12\catcode `\$12\catcode
  `\&12\catcode `\#12\catcode `\^12\catcode `\_12\catcode `\%12\relax}%
\providecommand \@@startlink[1]{}%
\providecommand \@@endlink[0]{}%
\providecommand \url  [0]{\begingroup\@sanitize@url \@url }%
\providecommand \@url [1]{\endgroup\@href {#1}{\urlprefix }}%
\providecommand \urlprefix  [0]{URL }%
\providecommand \Eprint [0]{\href }%
\providecommand \doibase [0]{https://doi.org/}%
\providecommand \selectlanguage [0]{\@gobble}%
\providecommand \bibinfo  [0]{\@secondoftwo}%
\providecommand \bibfield  [0]{\@secondoftwo}%
\providecommand \translation [1]{[#1]}%
\providecommand \BibitemOpen [0]{}%
\providecommand \bibitemStop [0]{}%
\providecommand \bibitemNoStop [0]{.\EOS\space}%
\providecommand \EOS [0]{\spacefactor3000\relax}%
\providecommand \BibitemShut  [1]{\csname bibitem#1\endcsname}%
\let\auto@bib@innerbib\@empty
\bibitem [{\citenamefont {Lamb}(1932)}]{Lamb:1932}%
  \BibitemOpen
  \bibfield  {author} {\bibinfo {author} {\bibfnamefont {H.}~\bibnamefont
  {Lamb}},\ }\href@noop {} {\emph {\bibinfo {title} {Hydrodynamics}}},\
  \bibinfo {edition} {6th}\ ed.\ (\bibinfo  {publisher} {Cambridge University
  press},\ \bibinfo {year} {1932})\BibitemShut {NoStop}%
\bibitem [{\citenamefont {Lugt}(1983)}]{Lugt1983}%
  \BibitemOpen
  \bibfield  {author} {\bibinfo {author} {\bibfnamefont {H.~J.}\ \bibnamefont
  {Lugt}},\ }\bibfield  {title} {\bibinfo {title} {Autorotation},\ }\href@noop
  {} {\bibfield  {journal} {\bibinfo  {journal} {Annu. Rev. Fluid Mech.}\
  }\textbf {\bibinfo {volume} {15}},\ \bibinfo {pages} {123} (\bibinfo {year}
  {1983})}\BibitemShut {NoStop}%
\bibitem [{\citenamefont {Bearman}(1984)}]{bearman1984}%
  \BibitemOpen
  \bibfield  {author} {\bibinfo {author} {\bibfnamefont {P.~W.}\ \bibnamefont
  {Bearman}},\ }\bibfield  {title} {\bibinfo {title} {Vortex shedding from
  oscillating bluff bodies},\ }\href@noop {} {\bibfield  {journal} {\bibinfo
  {journal} {Annu. Rev. Fluid Mech.}\ }\textbf {\bibinfo {volume} {16}},\
  \bibinfo {pages} {195} (\bibinfo {year} {1984})}\BibitemShut {NoStop}%
\bibitem [{\citenamefont {Williamson}\ and\ \citenamefont
  {Govardhan}(2004{\natexlab{a}})}]{williamson2004vortex}%
  \BibitemOpen
  \bibfield  {author} {\bibinfo {author} {\bibfnamefont {C.~H.~K.}\
  \bibnamefont {Williamson}}\ and\ \bibinfo {author} {\bibfnamefont
  {R.}~\bibnamefont {Govardhan}},\ }\bibfield  {title} {\bibinfo {title}
  {Vortex-induced vibrations},\ }\href@noop {} {\bibfield  {journal} {\bibinfo
  {journal} {Annu. Rev. Fluid Mech.}\ }\textbf {\bibinfo {volume} {36}},\
  \bibinfo {pages} {413} (\bibinfo {year} {2004}{\natexlab{a}})}\BibitemShut
  {NoStop}%
\bibitem [{\citenamefont {Ern}\ \emph {et~al.}(2012)\citenamefont {Ern},
  \citenamefont {Risso}, \citenamefont {Fabre},\ and\ \citenamefont
  {Magnaudet}}]{Ern:2012}%
  \BibitemOpen
  \bibfield  {author} {\bibinfo {author} {\bibfnamefont {P.}~\bibnamefont
  {Ern}}, \bibinfo {author} {\bibfnamefont {F.}~\bibnamefont {Risso}}, \bibinfo
  {author} {\bibfnamefont {D.}~\bibnamefont {Fabre}},\ and\ \bibinfo {author}
  {\bibfnamefont {J.}~\bibnamefont {Magnaudet}},\ }\bibfield  {title} {\bibinfo
  {title} {Wake-induced oscillatory paths of bodies freely rising or falling in
  fluids},\ }\href@noop {} {\bibfield  {journal} {\bibinfo  {journal} {Annu.
  Rev. Fluid Mech.}\ }\textbf {\bibinfo {volume} {44}},\ \bibinfo {pages} {97}
  (\bibinfo {year} {2012})}\BibitemShut {NoStop}%
\bibitem [{\citenamefont {Mathai}\ \emph {et~al.}(2020)\citenamefont {Mathai},
  \citenamefont {Lohse},\ and\ \citenamefont {Sun}}]{mathai2020}%
  \BibitemOpen
  \bibfield  {author} {\bibinfo {author} {\bibfnamefont {V.}~\bibnamefont
  {Mathai}}, \bibinfo {author} {\bibfnamefont {D.}~\bibnamefont {Lohse}},\ and\
  \bibinfo {author} {\bibfnamefont {C.}~\bibnamefont {Sun}},\ }\bibfield
  {title} {\bibinfo {title} {Bubbly and buoyant particle--laden turbulent
  flows},\ }\href@noop {} {\bibfield  {journal} {\bibinfo  {journal} {Annu.
  Rev. Condens. Matter Phys.}\ }\textbf {\bibinfo {volume} {11}},\ \bibinfo
  {pages} {529} (\bibinfo {year} {2020})}\BibitemShut {NoStop}%
\bibitem [{\citenamefont {Magnaudet}\ and\ \citenamefont
  {Eames}(2000)}]{Magnaudet2000}%
  \BibitemOpen
  \bibfield  {author} {\bibinfo {author} {\bibfnamefont {J.}~\bibnamefont
  {Magnaudet}}\ and\ \bibinfo {author} {\bibfnamefont {I.}~\bibnamefont
  {Eames}},\ }\bibfield  {title} {\bibinfo {title} {The motion of
  high-{R}eynolds-number bubbles in inhomogeneous flows},\ }\href@noop {}
  {\bibfield  {journal} {\bibinfo  {journal} {Annu. Rev. Fluid Mech.}\ }\textbf
  {\bibinfo {volume} {32}},\ \bibinfo {pages} {659} (\bibinfo {year}
  {2000})}\BibitemShut {NoStop}%
\bibitem [{\citenamefont {Zhao}\ \emph {et~al.}(2010)\citenamefont {Zhao},
  \citenamefont {Andersson},\ and\ \citenamefont {Gillissen}}]{zhao2010}%
  \BibitemOpen
  \bibfield  {author} {\bibinfo {author} {\bibfnamefont {L.~H.}\ \bibnamefont
  {Zhao}}, \bibinfo {author} {\bibfnamefont {H.~I.}\ \bibnamefont
  {Andersson}},\ and\ \bibinfo {author} {\bibfnamefont {J.~J.~J.}\ \bibnamefont
  {Gillissen}},\ }\bibfield  {title} {\bibinfo {title} {Turbulence modulation
  and drag reduction by spherical particles},\ }\href@noop {} {\bibfield
  {journal} {\bibinfo  {journal} {Phys. Fluids}\ }\textbf {\bibinfo {volume}
  {22}},\ \bibinfo {pages} {081702} (\bibinfo {year} {2010})}\BibitemShut
  {NoStop}%
\bibitem [{\citenamefont {Alm{\'e}ras}\ \emph {et~al.}(2015)\citenamefont
  {Alm{\'e}ras}, \citenamefont {Risso}, \citenamefont {Roig}, \citenamefont
  {Cazin}, \citenamefont {Plais},\ and\ \citenamefont {Augier}}]{almeras2015}%
  \BibitemOpen
  \bibfield  {author} {\bibinfo {author} {\bibfnamefont {E.}~\bibnamefont
  {Alm{\'e}ras}}, \bibinfo {author} {\bibfnamefont {F.}~\bibnamefont {Risso}},
  \bibinfo {author} {\bibfnamefont {V.}~\bibnamefont {Roig}}, \bibinfo {author}
  {\bibfnamefont {S.}~\bibnamefont {Cazin}}, \bibinfo {author} {\bibfnamefont
  {C.}~\bibnamefont {Plais}},\ and\ \bibinfo {author} {\bibfnamefont
  {F.}~\bibnamefont {Augier}},\ }\bibfield  {title} {\bibinfo {title} {Mixing
  by bubble-induced turbulence},\ }\href@noop {} {\bibfield  {journal}
  {\bibinfo  {journal} {J. Fluid Mech.}\ }\textbf {\bibinfo {volume} {776}},\
  \bibinfo {pages} {458} (\bibinfo {year} {2015})}\BibitemShut {NoStop}%
\bibitem [{\citenamefont {Alm{\'e}ras}\ \emph {et~al.}(2019)\citenamefont
  {Alm{\'e}ras}, \citenamefont {Mathai}, \citenamefont {Sun},\ and\
  \citenamefont {Lohse}}]{almeras2019}%
  \BibitemOpen
  \bibfield  {author} {\bibinfo {author} {\bibfnamefont {E.}~\bibnamefont
  {Alm{\'e}ras}}, \bibinfo {author} {\bibfnamefont {V.}~\bibnamefont {Mathai}},
  \bibinfo {author} {\bibfnamefont {C.}~\bibnamefont {Sun}},\ and\ \bibinfo
  {author} {\bibfnamefont {D.}~\bibnamefont {Lohse}},\ }\bibfield  {title}
  {\bibinfo {title} {Mixing induced by a bubble swarm rising through incident
  turbulence},\ }\href@noop {} {\bibfield  {journal} {\bibinfo  {journal} {Int.
  J. Multiph. Flow}\ }\textbf {\bibinfo {volume} {114}},\ \bibinfo {pages}
  {316} (\bibinfo {year} {2019})}\BibitemShut {NoStop}%
\bibitem [{\citenamefont {Auguste}\ and\ \citenamefont
  {Magnaudet}(2018)}]{Auguste2018}%
  \BibitemOpen
  \bibfield  {author} {\bibinfo {author} {\bibfnamefont {F.}~\bibnamefont
  {Auguste}}\ and\ \bibinfo {author} {\bibfnamefont {J.}~\bibnamefont
  {Magnaudet}},\ }\bibfield  {title} {\bibinfo {title} {Path oscillations and
  enhanced drag of light rising spheres},\ }\href@noop {} {\bibfield  {journal}
  {\bibinfo  {journal} {J. Fluid Mech.}\ }\textbf {\bibinfo {volume} {841}},\
  \bibinfo {pages} {228} (\bibinfo {year} {2018})}\BibitemShut {NoStop}%
\bibitem [{\citenamefont {Mathai}\ \emph {et~al.}(2018)\citenamefont {Mathai},
  \citenamefont {Zhu}, \citenamefont {Sun},\ and\ \citenamefont
  {Lohse}}]{Mathai2018}%
  \BibitemOpen
  \bibfield  {author} {\bibinfo {author} {\bibfnamefont {V.}~\bibnamefont
  {Mathai}}, \bibinfo {author} {\bibfnamefont {X.}~\bibnamefont {Zhu}},
  \bibinfo {author} {\bibfnamefont {C.}~\bibnamefont {Sun}},\ and\ \bibinfo
  {author} {\bibfnamefont {D.}~\bibnamefont {Lohse}},\ }\bibfield  {title}
  {\bibinfo {title} {Flutter to tumble transition of buoyant spheres triggered
  by rotational inertia changes},\ }\href@noop {} {\bibfield  {journal}
  {\bibinfo  {journal} {Nat. Comm.}\ }\textbf {\bibinfo {volume} {9}},\
  \bibinfo {pages} {1792} (\bibinfo {year} {2018})}\BibitemShut {NoStop}%
\bibitem [{\citenamefont {Jenny}\ \emph {et~al.}(2003)\citenamefont {Jenny},
  \citenamefont {Bouchet},\ and\ \citenamefont {Du\u{s}ek}}]{Jenny2003}%
  \BibitemOpen
  \bibfield  {author} {\bibinfo {author} {\bibfnamefont {M.}~\bibnamefont
  {Jenny}}, \bibinfo {author} {\bibfnamefont {G.}~\bibnamefont {Bouchet}},\
  and\ \bibinfo {author} {\bibfnamefont {J.}~\bibnamefont {Du\u{s}ek}},\
  }\bibfield  {title} {\bibinfo {title} {Nonvertical ascension or fall of a
  free sphere in a newtonian fluid},\ }\href@noop {} {\bibfield  {journal}
  {\bibinfo  {journal} {Phys. Fluids}\ }\textbf {\bibinfo {volume} {15}},\
  \bibinfo {pages} {L9} (\bibinfo {year} {2003})}\BibitemShut {NoStop}%
\bibitem [{\citenamefont {Horowitz}\ and\ \citenamefont
  {Williamson}(2008)}]{horowitz2008}%
  \BibitemOpen
  \bibfield  {author} {\bibinfo {author} {\bibfnamefont {M.}~\bibnamefont
  {Horowitz}}\ and\ \bibinfo {author} {\bibfnamefont {C.~H.~K.}\ \bibnamefont
  {Williamson}},\ }\bibfield  {title} {\bibinfo {title} {Critical mass and a
  new periodic four-ring vortex wake mode for freely rising and falling
  spheres},\ }\href@noop {} {\bibfield  {journal} {\bibinfo  {journal} {Phys.
  Fluids}\ }\textbf {\bibinfo {volume} {20}},\ \bibinfo {pages} {101701}
  (\bibinfo {year} {2008})}\BibitemShut {NoStop}%
\bibitem [{\citenamefont {Jenny}\ \emph {et~al.}(2004)\citenamefont {Jenny},
  \citenamefont {Du\u{s}ek},\ and\ \citenamefont {Bouchet}}]{Jenny2004}%
  \BibitemOpen
  \bibfield  {author} {\bibinfo {author} {\bibfnamefont {M.}~\bibnamefont
  {Jenny}}, \bibinfo {author} {\bibfnamefont {J.}~\bibnamefont {Du\u{s}ek}},\
  and\ \bibinfo {author} {\bibfnamefont {G.}~\bibnamefont {Bouchet}},\
  }\bibfield  {title} {\bibinfo {title} {Instabilities and transition of a
  sphere falling or ascending freely in a {N}ewtonian fluid},\ }\href@noop {}
  {\bibfield  {journal} {\bibinfo  {journal} {J. Fluid Mech.}\ }\textbf
  {\bibinfo {volume} {508}},\ \bibinfo {pages} {201} (\bibinfo {year}
  {2004})}\BibitemShut {NoStop}%
\bibitem [{\citenamefont {Veldhuis}\ \emph {et~al.}(2004)\citenamefont
  {Veldhuis}, \citenamefont {Biesheuvel}, \citenamefont {van Wijngaarden},\
  and\ \citenamefont {Lohse}}]{veldhuis2004}%
  \BibitemOpen
  \bibfield  {author} {\bibinfo {author} {\bibfnamefont {C.~H.~J.}\
  \bibnamefont {Veldhuis}}, \bibinfo {author} {\bibfnamefont {A.}~\bibnamefont
  {Biesheuvel}}, \bibinfo {author} {\bibfnamefont {L.}~\bibnamefont {van
  Wijngaarden}},\ and\ \bibinfo {author} {\bibfnamefont {D.}~\bibnamefont
  {Lohse}},\ }\bibfield  {title} {\bibinfo {title} {Motion and wake structure
  of spherical particles},\ }\href@noop {} {\bibfield  {journal} {\bibinfo
  {journal} {Nonlinearity}\ }\textbf {\bibinfo {volume} {18}},\ \bibinfo
  {pages} {C1} (\bibinfo {year} {2004})}\BibitemShut {NoStop}%
\bibitem [{\citenamefont {Veldhuis}\ and\ \citenamefont
  {Biesheuvel}(2007)}]{veldhuis2007}%
  \BibitemOpen
  \bibfield  {author} {\bibinfo {author} {\bibfnamefont {C.~H.~J.}\
  \bibnamefont {Veldhuis}}\ and\ \bibinfo {author} {\bibfnamefont
  {A.}~\bibnamefont {Biesheuvel}},\ }\bibfield  {title} {\bibinfo {title} {An
  experimental study of the regimes of motion of spheres falling or ascending
  freely in a {N}ewtonian fluid},\ }\href@noop {} {\bibfield  {journal}
  {\bibinfo  {journal} {Int. J. Multiph. Flow}\ }\textbf {\bibinfo {volume}
  {33}},\ \bibinfo {pages} {1074} (\bibinfo {year} {2007})}\BibitemShut
  {NoStop}%
\bibitem [{\citenamefont {Horowitz}\ and\ \citenamefont
  {Williamson}(2010)}]{Horowitz:2010}%
  \BibitemOpen
  \bibfield  {author} {\bibinfo {author} {\bibfnamefont {M.}~\bibnamefont
  {Horowitz}}\ and\ \bibinfo {author} {\bibfnamefont {C.~H.~K.}\ \bibnamefont
  {Williamson}},\ }\bibfield  {title} {\bibinfo {title} {The effect of
  {R}eynolds number on the dynamics and wakes of freely rising and falling
  spheres},\ }\href@noop {} {\bibfield  {journal} {\bibinfo  {journal} {J.
  Fluid Mech.}\ }\textbf {\bibinfo {volume} {651}},\ \bibinfo {pages} {251}
  (\bibinfo {year} {2010})}\BibitemShut {NoStop}%
\bibitem [{\citenamefont {Preukschat}(1962)}]{preukschat1962}%
  \BibitemOpen
  \bibfield  {author} {\bibinfo {author} {\bibfnamefont {A.~W.}\ \bibnamefont
  {Preukschat}},\ }\emph {\bibinfo {title} {Measurements of drag coefficients
  for falling and rising spheres in free motion}},\ \href@noop {} {Master's
  thesis},\ \bibinfo  {school} {California Institute of Technology} (\bibinfo
  {year} {1962})\BibitemShut {NoStop}%
\bibitem [{\citenamefont {Shafrir}(1965)}]{shafrir1965}%
  \BibitemOpen
  \bibfield  {author} {\bibinfo {author} {\bibfnamefont {U.}~\bibnamefont
  {Shafrir}},\ }\href@noop {} {\emph {\bibinfo {title} {Horizontal Oscillations
  of Falling Spheres}}},\ \bibinfo {type} {Tech. Rep.}\ (\bibinfo
  {institution} {Air Force Cambridge Research Labs.},\ \bibinfo {year}
  {1965})\BibitemShut {NoStop}%
\bibitem [{\citenamefont {Christiansen}\ and\ \citenamefont
  {Barker}(1965)}]{christiansen1965}%
  \BibitemOpen
  \bibfield  {author} {\bibinfo {author} {\bibfnamefont {E.~B.}\ \bibnamefont
  {Christiansen}}\ and\ \bibinfo {author} {\bibfnamefont {D.~H.}\ \bibnamefont
  {Barker}},\ }\bibfield  {title} {\bibinfo {title} {The effect of shape and
  density on the free settling of particles at high reynolds numbers},\
  }\href@noop {} {\bibfield  {journal} {\bibinfo  {journal} {AlChE J.}\
  }\textbf {\bibinfo {volume} {11}},\ \bibinfo {pages} {145} (\bibinfo {year}
  {1965})}\BibitemShut {NoStop}%
\bibitem [{\citenamefont {Karamanev}\ and\ \citenamefont
  {Nikolov}(1992)}]{Karamanev:1992}%
  \BibitemOpen
  \bibfield  {author} {\bibinfo {author} {\bibfnamefont {D.~G.}\ \bibnamefont
  {Karamanev}}\ and\ \bibinfo {author} {\bibfnamefont {L.~N.}\ \bibnamefont
  {Nikolov}},\ }\bibfield  {title} {\bibinfo {title} {Free rising spheres do
  not obey {N}ewton's law for free settling},\ }\href@noop {} {\bibfield
  {journal} {\bibinfo  {journal} {AIChE J.}\ }\textbf {\bibinfo {volume}
  {38}},\ \bibinfo {pages} {1843} (\bibinfo {year} {1992})}\BibitemShut
  {NoStop}%
\bibitem [{\citenamefont {Karamanev}\ \emph {et~al.}(1996)\citenamefont
  {Karamanev}, \citenamefont {Chavarie},\ and\ \citenamefont
  {Mayer}}]{karamanev1996}%
  \BibitemOpen
  \bibfield  {author} {\bibinfo {author} {\bibfnamefont {D.~G.}\ \bibnamefont
  {Karamanev}}, \bibinfo {author} {\bibfnamefont {C.}~\bibnamefont
  {Chavarie}},\ and\ \bibinfo {author} {\bibfnamefont {R.}~\bibnamefont
  {Mayer}},\ }\bibfield  {title} {\bibinfo {title} {Dynamics of the free rise
  of a light solid sphere in liquid},\ }\href@noop {} {\bibfield  {journal}
  {\bibinfo  {journal} {AIChE J.}\ }\textbf {\bibinfo {volume} {42}},\ \bibinfo
  {pages} {1789} (\bibinfo {year} {1996})}\BibitemShut {NoStop}%
\bibitem [{\citenamefont {Veldhuis}\ \emph {et~al.}(2009)\citenamefont
  {Veldhuis}, \citenamefont {Biesheuvel},\ and\ \citenamefont
  {Lohse}}]{veldhuis2009}%
  \BibitemOpen
  \bibfield  {author} {\bibinfo {author} {\bibfnamefont {C.~H.~J.}\
  \bibnamefont {Veldhuis}}, \bibinfo {author} {\bibfnamefont {A.}~\bibnamefont
  {Biesheuvel}},\ and\ \bibinfo {author} {\bibfnamefont {D.}~\bibnamefont
  {Lohse}},\ }\bibfield  {title} {\bibinfo {title} {Freely rising light solid
  spheres},\ }\href@noop {} {\bibfield  {journal} {\bibinfo  {journal} {Int. J.
  Multiph. Flow}\ }\textbf {\bibinfo {volume} {35}},\ \bibinfo {pages} {312}
  (\bibinfo {year} {2009})}\BibitemShut {NoStop}%
\bibitem [{\citenamefont {Namkoong}\ \emph {et~al.}(2008)\citenamefont
  {Namkoong}, \citenamefont {Yoo},\ and\ \citenamefont {Choi}}]{namkoong2008}%
  \BibitemOpen
  \bibfield  {author} {\bibinfo {author} {\bibfnamefont {K.}~\bibnamefont
  {Namkoong}}, \bibinfo {author} {\bibfnamefont {J.~Y.}\ \bibnamefont {Yoo}},\
  and\ \bibinfo {author} {\bibfnamefont {H.~G.}\ \bibnamefont {Choi}},\
  }\bibfield  {title} {\bibinfo {title} {Numerical analysis of two-dimensional
  motion of a freely falling circular cylinder in an infinite fluid},\
  }\href@noop {} {\bibfield  {journal} {\bibinfo  {journal} {J. Fluid Mech.}\
  }\textbf {\bibinfo {volume} {604}},\ \bibinfo {pages} {33} (\bibinfo {year}
  {2008})}\BibitemShut {NoStop}%
\bibitem [{\citenamefont {Mathai}\ \emph {et~al.}(2017)\citenamefont {Mathai},
  \citenamefont {Zhu}, \citenamefont {Sun},\ and\ \citenamefont
  {Lohse}}]{Mathai:2017}%
  \BibitemOpen
  \bibfield  {author} {\bibinfo {author} {\bibfnamefont {V.}~\bibnamefont
  {Mathai}}, \bibinfo {author} {\bibfnamefont {X.}~\bibnamefont {Zhu}},
  \bibinfo {author} {\bibfnamefont {C.}~\bibnamefont {Sun}},\ and\ \bibinfo
  {author} {\bibfnamefont {D.}~\bibnamefont {Lohse}},\ }\bibfield  {title}
  {\bibinfo {title} {Mass and moment of inertia govern the transition in the
  dynamics and wakes of freely rising and falling cylinders.},\ }\href@noop {}
  {\bibfield  {journal} {\bibinfo  {journal} {Phys. Rev. Lett.}\ }\textbf
  {\bibinfo {volume} {119}},\ \bibinfo {pages} {054501} (\bibinfo {year}
  {2017})}\BibitemShut {NoStop}%
\bibitem [{\citenamefont {Loth}(2008)}]{loth2008lift}%
  \BibitemOpen
  \bibfield  {author} {\bibinfo {author} {\bibfnamefont {E.}~\bibnamefont
  {Loth}},\ }\bibfield  {title} {\bibinfo {title} {Lift of a spherical particle
  subject to vorticity and/or spin},\ }\href@noop {} {\bibfield  {journal}
  {\bibinfo  {journal} {AIAA J.}\ }\textbf {\bibinfo {volume} {46}},\ \bibinfo
  {pages} {801} (\bibinfo {year} {2008})}\BibitemShut {NoStop}%
\bibitem [{\citenamefont {Cummins}\ \emph {et~al.}(2018)\citenamefont
  {Cummins}, \citenamefont {Seale}, \citenamefont {Macente}, \citenamefont
  {Certini}, \citenamefont {Mastropaolo}, \citenamefont {Viola},\ and\
  \citenamefont {Nakayama}}]{cummins2018}%
  \BibitemOpen
  \bibfield  {author} {\bibinfo {author} {\bibfnamefont {C.}~\bibnamefont
  {Cummins}}, \bibinfo {author} {\bibfnamefont {M.}~\bibnamefont {Seale}},
  \bibinfo {author} {\bibfnamefont {A.}~\bibnamefont {Macente}}, \bibinfo
  {author} {\bibfnamefont {D.}~\bibnamefont {Certini}}, \bibinfo {author}
  {\bibfnamefont {E.}~\bibnamefont {Mastropaolo}}, \bibinfo {author}
  {\bibfnamefont {I.~M.}\ \bibnamefont {Viola}},\ and\ \bibinfo {author}
  {\bibfnamefont {N.}~\bibnamefont {Nakayama}},\ }\bibfield  {title} {\bibinfo
  {title} {A separated vortex ring underlies the flight of the dandelion},\
  }\href@noop {} {\bibfield  {journal} {\bibinfo  {journal} {Nature}\ }\textbf
  {\bibinfo {volume} {562}},\ \bibinfo {pages} {414} (\bibinfo {year}
  {2018})}\BibitemShut {NoStop}%
\bibitem [{\citenamefont {Nemes}\ \emph {et~al.}(2017)\citenamefont {Nemes},
  \citenamefont {Dasari}, \citenamefont {Hong}, \citenamefont {Guala},\ and\
  \citenamefont {Coletti}}]{nemes2017}%
  \BibitemOpen
  \bibfield  {author} {\bibinfo {author} {\bibfnamefont {A.}~\bibnamefont
  {Nemes}}, \bibinfo {author} {\bibfnamefont {T.}~\bibnamefont {Dasari}},
  \bibinfo {author} {\bibfnamefont {J.}~\bibnamefont {Hong}}, \bibinfo {author}
  {\bibfnamefont {M.}~\bibnamefont {Guala}},\ and\ \bibinfo {author}
  {\bibfnamefont {F.}~\bibnamefont {Coletti}},\ }\bibfield  {title} {\bibinfo
  {title} {Snowflakes in the atmospheric surface layer: observation of
  particle--turbulence dynamics},\ }\href@noop {} {\bibfield  {journal}
  {\bibinfo  {journal} {J. Fluid Mech.}\ }\textbf {\bibinfo {volume} {814}},\
  \bibinfo {pages} {592} (\bibinfo {year} {2017})}\BibitemShut {NoStop}%
\bibitem [{\citenamefont {Li}\ \emph {et~al.}(2020)\citenamefont {Li},
  \citenamefont {Lim}, \citenamefont {Berk}, \citenamefont {Abraham},
  \citenamefont {Heisel}, \citenamefont {Guala}, \citenamefont {Coletti},\ and\
  \citenamefont {Hong}}]{li2020}%
  \BibitemOpen
  \bibfield  {author} {\bibinfo {author} {\bibfnamefont {C.}~\bibnamefont
  {Li}}, \bibinfo {author} {\bibfnamefont {K.}~\bibnamefont {Lim}}, \bibinfo
  {author} {\bibfnamefont {T.}~\bibnamefont {Berk}}, \bibinfo {author}
  {\bibfnamefont {A.}~\bibnamefont {Abraham}}, \bibinfo {author} {\bibfnamefont
  {M.}~\bibnamefont {Heisel}}, \bibinfo {author} {\bibfnamefont
  {M.}~\bibnamefont {Guala}}, \bibinfo {author} {\bibfnamefont
  {F.}~\bibnamefont {Coletti}},\ and\ \bibinfo {author} {\bibfnamefont
  {J.}~\bibnamefont {Hong}},\ }\bibfield  {title} {\bibinfo {title} {Settling
  and clustering of snow particles in atmospheric turbulence},\ }\href@noop {}
  {\bibfield  {journal} {\bibinfo  {journal} {arXiv preprint arXiv:2006.09502}\
  } (\bibinfo {year} {2020})}\BibitemShut {NoStop}%
\bibitem [{\citenamefont {Zeugin}\ \emph {et~al.}(2020)\citenamefont {Zeugin},
  \citenamefont {Krol}, \citenamefont {Fouxon},\ and\ \citenamefont
  {Holzner}}]{zeugin2020}%
  \BibitemOpen
  \bibfield  {author} {\bibinfo {author} {\bibfnamefont {T.}~\bibnamefont
  {Zeugin}}, \bibinfo {author} {\bibfnamefont {Q.}~\bibnamefont {Krol}},
  \bibinfo {author} {\bibfnamefont {I.}~\bibnamefont {Fouxon}},\ and\ \bibinfo
  {author} {\bibfnamefont {M.}~\bibnamefont {Holzner}},\ }\bibfield  {title}
  {\bibinfo {title} {Sedimentation of snow particles in still air in stokes
  regime},\ }\href@noop {} {\bibfield  {journal} {\bibinfo  {journal} {Geophys.
  Res. Lett.}\ }\textbf {\bibinfo {volume} {47}},\ \bibinfo {pages}
  {e2020GL087832} (\bibinfo {year} {2020})}\BibitemShut {NoStop}%
\bibitem [{\citenamefont {McCorquodale}\ and\ \citenamefont
  {Westbrook}(2020{\natexlab{a}})}]{mccorquodaletrail_p1}%
  \BibitemOpen
  \bibfield  {author} {\bibinfo {author} {\bibfnamefont {M.~W.}\ \bibnamefont
  {McCorquodale}}\ and\ \bibinfo {author} {\bibfnamefont {C.}~\bibnamefont
  {Westbrook}},\ }\bibfield  {title} {\bibinfo {title} {Trail: a novel approach
  for studying the aerodynamics of ice particles},\ }\href@noop {} {\bibfield
  {journal} {\bibinfo  {journal} {Q. J. R. Meteorol. Soc.}\ } (\bibinfo {year}
  {2020}{\natexlab{a}})}\BibitemShut {NoStop}%
\bibitem [{\citenamefont {McCorquodale}\ and\ \citenamefont
  {Westbrook}(2020{\natexlab{b}})}]{mccorquodaletrail_p2}%
  \BibitemOpen
  \bibfield  {author} {\bibinfo {author} {\bibfnamefont {M.~W.}\ \bibnamefont
  {McCorquodale}}\ and\ \bibinfo {author} {\bibfnamefont {C.}~\bibnamefont
  {Westbrook}},\ }\bibfield  {title} {\bibinfo {title} {Trail part 2: a
  comprehensive assessment of ice particle fall speed parametrisations},\
  }\href@noop {} {\bibfield  {journal} {\bibinfo  {journal} {Q. J. R. Meteorol.
  Soc.}\ } (\bibinfo {year} {2020}{\natexlab{b}})}\BibitemShut {NoStop}%
\bibitem [{\citenamefont {Richardson}\ and\ \citenamefont
  {Zaki}(1954)}]{richardson1954}%
  \BibitemOpen
  \bibfield  {author} {\bibinfo {author} {\bibfnamefont {J.~F.}\ \bibnamefont
  {Richardson}}\ and\ \bibinfo {author} {\bibfnamefont {W.~N.}\ \bibnamefont
  {Zaki}},\ }\bibfield  {title} {\bibinfo {title} {The sedimentation of a
  suspension of uniform spheres under conditions of viscous flow},\ }\href@noop
  {} {\bibfield  {journal} {\bibinfo  {journal} {Chem. Eng. Sci.}\ }\textbf
  {\bibinfo {volume} {3}},\ \bibinfo {pages} {65} (\bibinfo {year}
  {1954})}\BibitemShut {NoStop}%
\bibitem [{\citenamefont {Meiburg}\ and\ \citenamefont
  {Kneller}(2010)}]{meiburg2010}%
  \BibitemOpen
  \bibfield  {author} {\bibinfo {author} {\bibfnamefont {E.}~\bibnamefont
  {Meiburg}}\ and\ \bibinfo {author} {\bibfnamefont {B.}~\bibnamefont
  {Kneller}},\ }\bibfield  {title} {\bibinfo {title} {Turbidity currents and
  their deposits},\ }\href@noop {} {\bibfield  {journal} {\bibinfo  {journal}
  {Annu. Rev. Fluid Mech.}\ }\textbf {\bibinfo {volume} {42}},\ \bibinfo
  {pages} {135} (\bibinfo {year} {2010})}\BibitemShut {NoStop}%
\bibitem [{\citenamefont {Sowmeyan}\ and\ \citenamefont
  {Swaminathan}(2008)}]{sowmeyan2008}%
  \BibitemOpen
  \bibfield  {author} {\bibinfo {author} {\bibfnamefont {R.}~\bibnamefont
  {Sowmeyan}}\ and\ \bibinfo {author} {\bibfnamefont {G.}~\bibnamefont
  {Swaminathan}},\ }\bibfield  {title} {\bibinfo {title} {Evaluation of inverse
  anaerobic fluidized bed reactor for treating high strength organic
  wastewater},\ }\href@noop {} {\bibfield  {journal} {\bibinfo  {journal}
  {Bioresour. Technol.}\ }\textbf {\bibinfo {volume} {99}},\ \bibinfo {pages}
  {3877} (\bibinfo {year} {2008})}\BibitemShut {NoStop}%
\bibitem [{\citenamefont {Clark}\ \emph {et~al.}(2020)\citenamefont {Clark},
  \citenamefont {DiBenedetto}, \citenamefont {Ouellette},\ and\ \citenamefont
  {Koseff}}]{clark2020settling}%
  \BibitemOpen
  \bibfield  {author} {\bibinfo {author} {\bibfnamefont {L.~K.}\ \bibnamefont
  {Clark}}, \bibinfo {author} {\bibfnamefont {M.~H.}\ \bibnamefont
  {DiBenedetto}}, \bibinfo {author} {\bibfnamefont {N.~T.}\ \bibnamefont
  {Ouellette}},\ and\ \bibinfo {author} {\bibfnamefont {J.~R.}\ \bibnamefont
  {Koseff}},\ }\bibfield  {title} {\bibinfo {title} {Settling of inertial
  nonspherical particles in wavy flow},\ }\href
  {https://doi.org/10.1103/PhysRevFluids.5.124301} {\bibfield  {journal}
  {\bibinfo  {journal} {Phys. Rev. Fluids}\ }\textbf {\bibinfo {volume} {5}},\
  \bibinfo {pages} {124301} (\bibinfo {year} {2020})}\BibitemShut {NoStop}%
\bibitem [{\citenamefont {Tanaka}\ \emph {et~al.}(2020)\citenamefont {Tanaka},
  \citenamefont {Tajiri}, \citenamefont {Nishida},\ and\ \citenamefont
  {Yamakawa}}]{tanaka2020}%
  \BibitemOpen
  \bibfield  {author} {\bibinfo {author} {\bibfnamefont {M.}~\bibnamefont
  {Tanaka}}, \bibinfo {author} {\bibfnamefont {K.}~\bibnamefont {Tajiri}},
  \bibinfo {author} {\bibfnamefont {H.}~\bibnamefont {Nishida}},\ and\ \bibinfo
  {author} {\bibfnamefont {M.}~\bibnamefont {Yamakawa}},\ }\bibfield  {title}
  {\bibinfo {title} {Effect of eccentric mass distribution on the motion of
  spherical particles in shear flows},\ }\href@noop {} {\bibfield  {journal}
  {\bibinfo  {journal} {J. Fluids Eng.}\ }\textbf {\bibinfo {volume} {142}},\
  \bibinfo {pages} {031105} (\bibinfo {year} {2020})}\BibitemShut {NoStop}%
\bibitem [{\citenamefont {Yasseri}(2014)}]{yasseri2014}%
  \BibitemOpen
  \bibfield  {author} {\bibinfo {author} {\bibfnamefont {S.}~\bibnamefont
  {Yasseri}},\ }\bibfield  {title} {\bibinfo {title} {Experiment of
  free-falling cylinders in water},\ }\href@noop {} {\bibfield  {journal}
  {\bibinfo  {journal} {Underw. Technol.}\ }\textbf {\bibinfo {volume} {32}},\
  \bibinfo {pages} {177} (\bibinfo {year} {2014})}\BibitemShut {NoStop}%
\bibitem [{\citenamefont {Angle}\ \emph {et~al.}(2019)\citenamefont {Angle},
  \citenamefont {Rau},\ and\ \citenamefont {Byron}}]{angle2019}%
  \BibitemOpen
  \bibfield  {author} {\bibinfo {author} {\bibfnamefont {B.~R.}\ \bibnamefont
  {Angle}}, \bibinfo {author} {\bibfnamefont {M.~J.}\ \bibnamefont {Rau}},\
  and\ \bibinfo {author} {\bibfnamefont {M.~L.}\ \bibnamefont {Byron}},\
  }\bibfield  {title} {\bibinfo {title} {Effect of mass distribution on falling
  cylindrical particles at intermediate {R}eynolds numbers},\ }in\ \href@noop
  {} {\emph {\bibinfo {booktitle} {ASME 2019 Fluids Engineering Division Summer
  Meeting}}}\ (\bibinfo {organization} {American Society of Mechanical
  Engineers},\ \bibinfo {year} {2019})\BibitemShut {NoStop}%
\bibitem [{\citenamefont {Roy}\ \emph {et~al.}(2019)\citenamefont {Roy},
  \citenamefont {Hamati}, \citenamefont {Tierney}, \citenamefont {Koch},\ and\
  \citenamefont {Voth}}]{roy2019}%
  \BibitemOpen
  \bibfield  {author} {\bibinfo {author} {\bibfnamefont {A.}~\bibnamefont
  {Roy}}, \bibinfo {author} {\bibfnamefont {R.~J.}\ \bibnamefont {Hamati}},
  \bibinfo {author} {\bibfnamefont {L.}~\bibnamefont {Tierney}}, \bibinfo
  {author} {\bibfnamefont {D.~L.}\ \bibnamefont {Koch}},\ and\ \bibinfo
  {author} {\bibfnamefont {G.~A.}\ \bibnamefont {Voth}},\ }\bibfield  {title}
  {\bibinfo {title} {Inertial torques and a symmetry breaking orientational
  transition in the sedimentation of slender fibres},\ }\href@noop {}
  {\bibfield  {journal} {\bibinfo  {journal} {J. Fluid Mech.}\ }\textbf
  {\bibinfo {volume} {875}},\ \bibinfo {pages} {576} (\bibinfo {year}
  {2019})}\BibitemShut {NoStop}%
\bibitem [{\citenamefont {Mougin}\ and\ \citenamefont
  {Magnaudet}(2002)}]{mougin2002}%
  \BibitemOpen
  \bibfield  {author} {\bibinfo {author} {\bibfnamefont {G.}~\bibnamefont
  {Mougin}}\ and\ \bibinfo {author} {\bibfnamefont {J.}~\bibnamefont
  {Magnaudet}},\ }\bibfield  {title} {\bibinfo {title} {The generalized
  {K}irchhoff equations and their application to the interaction between a
  rigid body and an arbitrary time-dependent viscous flow},\ }\href@noop {}
  {\bibfield  {journal} {\bibinfo  {journal} {Int. J. Multiph. Flow}\ }\textbf
  {\bibinfo {volume} {28}},\ \bibinfo {pages} {1837} (\bibinfo {year}
  {2002})}\BibitemShut {NoStop}%
\bibitem [{\citenamefont {Williamson}\ and\ \citenamefont
  {Govardhan}(2004{\natexlab{b}})}]{govardhan2004}%
  \BibitemOpen
  \bibfield  {author} {\bibinfo {author} {\bibfnamefont {C.~H.~K.}\
  \bibnamefont {Williamson}}\ and\ \bibinfo {author} {\bibfnamefont
  {R.}~\bibnamefont {Govardhan}},\ }\bibfield  {title} {\bibinfo {title}
  {Vortex-induced vibrations},\ }\href
  {https://doi.org/10.1146/annurev.fluid.36.050802.122128} {\bibfield
  {journal} {\bibinfo  {journal} {Annu. Rev. Fluid Mech.}\ }\textbf {\bibinfo
  {volume} {36}},\ \bibinfo {pages} {413} (\bibinfo {year}
  {2004}{\natexlab{b}})}\BibitemShut {NoStop}%
\bibitem [{\citenamefont {Belmonte}\ \emph {et~al.}(1998)\citenamefont
  {Belmonte}, \citenamefont {Eisenberg},\ and\ \citenamefont
  {Moses}}]{Belmonte:1998}%
  \BibitemOpen
  \bibfield  {author} {\bibinfo {author} {\bibfnamefont {A.}~\bibnamefont
  {Belmonte}}, \bibinfo {author} {\bibfnamefont {H.}~\bibnamefont
  {Eisenberg}},\ and\ \bibinfo {author} {\bibfnamefont {E.}~\bibnamefont
  {Moses}},\ }\bibfield  {title} {\bibinfo {title} {From flutter to tumble:
  Inertial drag and froude similarity in falling paper},\ }\href@noop {}
  {\bibfield  {journal} {\bibinfo  {journal} {Phys. Rev. Lett.}\ }\textbf
  {\bibinfo {volume} {81}},\ \bibinfo {pages} {345} (\bibinfo {year}
  {1998})}\BibitemShut {NoStop}%
\bibitem [{\citenamefont {Mathai}\ \emph {et~al.}(2016)\citenamefont {Mathai},
  \citenamefont {Neut}, \citenamefont {van~der Poel},\ and\ \citenamefont
  {Sun}}]{Mathai:2016}%
  \BibitemOpen
  \bibfield  {author} {\bibinfo {author} {\bibfnamefont {V.}~\bibnamefont
  {Mathai}}, \bibinfo {author} {\bibfnamefont {M.~W.~M.}\ \bibnamefont {Neut}},
  \bibinfo {author} {\bibfnamefont {E.~P.}\ \bibnamefont {van~der Poel}},\ and\
  \bibinfo {author} {\bibfnamefont {C.}~\bibnamefont {Sun}},\ }\bibfield
  {title} {\bibinfo {title} {Translational and rotational dynamics of a large
  buoyant sphere in turbulence},\ }\href@noop {} {\bibfield  {journal}
  {\bibinfo  {journal} {Exp. Fluids}\ }\textbf {\bibinfo {volume} {57}}
  (\bibinfo {year} {2016})}\BibitemShut {NoStop}%
\bibitem [{\citenamefont {Will}\ \emph {et~al.}(2020)\citenamefont {Will},
  \citenamefont {Mathai}, \citenamefont {Huisman}, \citenamefont {Lohse},
  \citenamefont {Sun},\ and\ \citenamefont {Krug}}]{will2020kinematics}%
  \BibitemOpen
  \bibfield  {author} {\bibinfo {author} {\bibfnamefont {J.~B.}\ \bibnamefont
  {Will}}, \bibinfo {author} {\bibfnamefont {V.}~\bibnamefont {Mathai}},
  \bibinfo {author} {\bibfnamefont {S.~G.}\ \bibnamefont {Huisman}}, \bibinfo
  {author} {\bibfnamefont {D.}~\bibnamefont {Lohse}}, \bibinfo {author}
  {\bibfnamefont {C.}~\bibnamefont {Sun}},\ and\ \bibinfo {author}
  {\bibfnamefont {D.}~\bibnamefont {Krug}},\ }\bibfield  {title} {\bibinfo
  {title} {Kinematics and dynamics of freely rising ellipsoids at high
  {R}eynolds numbers},\ }\href@noop {} {\bibfield  {journal} {\bibinfo
  {journal} {arXiv preprint arXiv:2007.06228}\ } (\bibinfo {year}
  {2020})}\BibitemShut {NoStop}%
\bibitem [{\citenamefont {Bishop}\ and\ \citenamefont
  {Hassan}(1964)}]{bishop1964lift}%
  \BibitemOpen
  \bibfield  {author} {\bibinfo {author} {\bibfnamefont {R.~E.~D.}\
  \bibnamefont {Bishop}}\ and\ \bibinfo {author} {\bibfnamefont {A.~Y.}\
  \bibnamefont {Hassan}},\ }\bibfield  {title} {\bibinfo {title} {The lift and
  drag forces on a circular cylinder oscillating in a flowing fluid},\
  }\href@noop {} {\bibfield  {journal} {\bibinfo  {journal} {Proc. R. Soc.
  Lond. Ser. A}\ }\textbf {\bibinfo {volume} {277}},\ \bibinfo {pages} {51}
  (\bibinfo {year} {1964})}\BibitemShut {NoStop}%
\bibitem [{\citenamefont {Bearman}\ and\ \citenamefont
  {Obasaju}(1982)}]{bearman1982experimental}%
  \BibitemOpen
  \bibfield  {author} {\bibinfo {author} {\bibfnamefont {P.~W.}\ \bibnamefont
  {Bearman}}\ and\ \bibinfo {author} {\bibfnamefont {E.~D.}\ \bibnamefont
  {Obasaju}},\ }\bibfield  {title} {\bibinfo {title} {An experimental study of
  pressure fluctuations on fixed and oscillating square-section cylinders},\
  }\href@noop {} {\bibfield  {journal} {\bibinfo  {journal} {J. Fluid Mech.}\
  }\textbf {\bibinfo {volume} {119}},\ \bibinfo {pages} {297} (\bibinfo {year}
  {1982})}\BibitemShut {NoStop}%
\bibitem [{\citenamefont {Jordan}\ and\ \citenamefont
  {Fromm}(1972)}]{jordan1972}%
  \BibitemOpen
  \bibfield  {author} {\bibinfo {author} {\bibfnamefont {S.~K.}\ \bibnamefont
  {Jordan}}\ and\ \bibinfo {author} {\bibfnamefont {J.~E.}\ \bibnamefont
  {Fromm}},\ }\bibfield  {title} {\bibinfo {title} {Oscillatory drag, lift, and
  torque on a circular cylinder in a uniform flow},\ }\href@noop {} {\bibfield
  {journal} {\bibinfo  {journal} {Phys. Fluids}\ }\textbf {\bibinfo {volume}
  {15}},\ \bibinfo {pages} {371} (\bibinfo {year} {1972})}\BibitemShut
  {NoStop}%
\bibitem [{\citenamefont {Bouchet}\ \emph {et~al.}(2006)\citenamefont
  {Bouchet}, \citenamefont {Mebarek},\ and\ \citenamefont
  {Du\u{s}ek}}]{bouchet2006}%
  \BibitemOpen
  \bibfield  {author} {\bibinfo {author} {\bibfnamefont {G.}~\bibnamefont
  {Bouchet}}, \bibinfo {author} {\bibfnamefont {M.}~\bibnamefont {Mebarek}},\
  and\ \bibinfo {author} {\bibfnamefont {J.}~\bibnamefont {Du\u{s}ek}},\
  }\bibfield  {title} {\bibinfo {title} {Hydrodynamic forces acting on a rigid
  fixed sphere in early transitional regimes},\ }\href@noop {} {\bibfield
  {journal} {\bibinfo  {journal} {Eur. J. Mech. B/Fluids}\ }\textbf {\bibinfo
  {volume} {25}},\ \bibinfo {pages} {321} (\bibinfo {year} {2006})}\BibitemShut
  {NoStop}%
\bibitem [{\citenamefont {Zimmermann}\ \emph {et~al.}(2011)\citenamefont
  {Zimmermann}, \citenamefont {Gasteuil}, \citenamefont {Bourgoin},
  \citenamefont {Volk}, \citenamefont {Pumir},\ and\ \citenamefont
  {Pinton}}]{zimmermann2011rotational}%
  \BibitemOpen
  \bibfield  {author} {\bibinfo {author} {\bibfnamefont {R.}~\bibnamefont
  {Zimmermann}}, \bibinfo {author} {\bibfnamefont {Y.}~\bibnamefont
  {Gasteuil}}, \bibinfo {author} {\bibfnamefont {M.}~\bibnamefont {Bourgoin}},
  \bibinfo {author} {\bibfnamefont {R.}~\bibnamefont {Volk}}, \bibinfo {author}
  {\bibfnamefont {A.}~\bibnamefont {Pumir}},\ and\ \bibinfo {author}
  {\bibfnamefont {J.-F.}\ \bibnamefont {Pinton}},\ }\bibfield  {title}
  {\bibinfo {title} {Rotational intermittency and turbulence induced lift
  experienced by large particles in a turbulent flow},\ }\href@noop {}
  {\bibfield  {journal} {\bibinfo  {journal} {Phys. Rev. Lett.}\ }\textbf
  {\bibinfo {volume} {106}},\ \bibinfo {pages} {154501} (\bibinfo {year}
  {2011})}\BibitemShut {NoStop}%
\bibitem [{\citenamefont {MacCready}\ and\ \citenamefont
  {Jex}(1964)}]{maccready1964}%
  \BibitemOpen
  \bibfield  {author} {\bibinfo {author} {\bibfnamefont {P.~B.}\ \bibnamefont
  {MacCready}}\ and\ \bibinfo {author} {\bibfnamefont {H.~R.}\ \bibnamefont
  {Jex}},\ }\href@noop {} {\emph {\bibinfo {title} {Study of sphere motion and
  balloon wind sensors}}},\ \bibinfo {type} {Tech. Rep.}\ (\bibinfo
  {institution} {NASA TM X-53089},\ \bibinfo {year} {1964})\BibitemShut
  {NoStop}%
\bibitem [{\citenamefont {Kuwabara}\ \emph {et~al.}(1983)\citenamefont
  {Kuwabara}, \citenamefont {Chiba},\ and\ \citenamefont
  {Kono}}]{kuwabara1983}%
  \BibitemOpen
  \bibfield  {author} {\bibinfo {author} {\bibfnamefont {G.}~\bibnamefont
  {Kuwabara}}, \bibinfo {author} {\bibfnamefont {S.}~\bibnamefont {Chiba}},\
  and\ \bibinfo {author} {\bibfnamefont {K.}~\bibnamefont {Kono}},\ }\bibfield
  {title} {\bibinfo {title} {Anomalous motion of a sphere falling through
  water},\ }\href@noop {} {\bibfield  {journal} {\bibinfo  {journal} {J. Phys.
  Soc. Jpn.}\ }\textbf {\bibinfo {volume} {52}},\ \bibinfo {pages} {3373}
  (\bibinfo {year} {1983})}\BibitemShut {NoStop}%
\bibitem [{\citenamefont {Stringham}\ \emph {et~al.}(1969)\citenamefont
  {Stringham}, \citenamefont {Simons},\ and\ \citenamefont
  {Guy}}]{stringham1969}%
  \BibitemOpen
  \bibfield  {author} {\bibinfo {author} {\bibfnamefont {G.~E.}\ \bibnamefont
  {Stringham}}, \bibinfo {author} {\bibfnamefont {D.~B.}\ \bibnamefont
  {Simons}},\ and\ \bibinfo {author} {\bibfnamefont {H.~P.}\ \bibnamefont
  {Guy}},\ }\bibfield  {title} {\bibinfo {title} {The behavior of large
  particles falling in quiescent liquids},\ }\href@noop {} {\bibfield
  {journal} {\bibinfo  {journal} {Prof. Pap. US Geol. Surv.}\ }\textbf
  {\bibinfo {volume} {562-C}} (\bibinfo {year} {1969})}\BibitemShut {NoStop}%
\bibitem [{\citenamefont {Allen}(1900)}]{allen1900}%
  \BibitemOpen
  \bibfield  {author} {\bibinfo {author} {\bibfnamefont {H.~S.}\ \bibnamefont
  {Allen}},\ }\bibfield  {title} {\bibinfo {title} {The motion of a sphere in a
  viscous fluid: {III}},\ }\href@noop {} {\bibfield  {journal} {\bibinfo
  {journal} {Phil. Mag.}\ }\textbf {\bibinfo {volume} {50}},\ \bibinfo {pages}
  {519} (\bibinfo {year} {1900})}\BibitemShut {NoStop}%
\bibitem [{\citenamefont {Liebster}(1927)}]{liebster1927}%
  \BibitemOpen
  \bibfield  {author} {\bibinfo {author} {\bibfnamefont {H.}~\bibnamefont
  {Liebster}},\ }\bibfield  {title} {\bibinfo {title} {{\"U}ber den widerstand
  von kugeln},\ }\href@noop {} {\bibfield  {journal} {\bibinfo  {journal} {Ann.
  Phys.}\ }\textbf {\bibinfo {volume} {387}},\ \bibinfo {pages} {541} (\bibinfo
  {year} {1927})}\BibitemShut {NoStop}%
\bibitem [{\citenamefont {Lunnon}(1928)}]{lunnon1928}%
  \BibitemOpen
  \bibfield  {author} {\bibinfo {author} {\bibfnamefont {R.~G.}\ \bibnamefont
  {Lunnon}},\ }\bibfield  {title} {\bibinfo {title} {Fluid resistance to moving
  spheres},\ }\href@noop {} {\bibfield  {journal} {\bibinfo  {journal} {Proc.
  R. Soc. Lond. A}\ }\textbf {\bibinfo {volume} {118}},\ \bibinfo {pages} {680}
  (\bibinfo {year} {1928})}\BibitemShut {NoStop}%
\bibitem [{\citenamefont {Boillat}\ and\ \citenamefont
  {Graf}(1981)}]{boillat1981}%
  \BibitemOpen
  \bibfield  {author} {\bibinfo {author} {\bibfnamefont {J.~L.}\ \bibnamefont
  {Boillat}}\ and\ \bibinfo {author} {\bibfnamefont {W.~H.}\ \bibnamefont
  {Graf}},\ }\bibfield  {title} {\bibinfo {title} {Settling velocity of
  spherical particles in calm water},\ }\href@noop {} {\bibfield  {journal}
  {\bibinfo  {journal} {J. Hydraul. Div.}\ }\textbf {\bibinfo {volume} {107}},\
  \bibinfo {pages} {1123} (\bibinfo {year} {1981})}\BibitemShut {NoStop}%
\bibitem [{\citenamefont {Newton}(1999)}]{newton1999}%
  \BibitemOpen
  \bibfield  {author} {\bibinfo {author} {\bibfnamefont {I.}~\bibnamefont
  {Newton}},\ }\href@noop {} {\emph {\bibinfo {title} {The Principia:
  mathematical principles of natural philosophy}}}\ (\bibinfo  {publisher}
  {Univ. of California Press},\ \bibinfo {year} {1999})\BibitemShut {NoStop}%
\bibitem [{\citenamefont {Zhou}\ and\ \citenamefont
  {Du{\v{s}}ek}(2015)}]{zhou2015chaotic}%
  \BibitemOpen
  \bibfield  {author} {\bibinfo {author} {\bibfnamefont {W.}~\bibnamefont
  {Zhou}}\ and\ \bibinfo {author} {\bibfnamefont {J.}~\bibnamefont
  {Du{\v{s}}ek}},\ }\bibfield  {title} {\bibinfo {title} {Chaotic states and
  order in the chaos of the paths of freely falling and ascending spheres},\
  }\href@noop {} {\bibfield  {journal} {\bibinfo  {journal} {Int. J. Multiph.
  Flow}\ }\textbf {\bibinfo {volume} {75}},\ \bibinfo {pages} {205} (\bibinfo
  {year} {2015})}\BibitemShut {NoStop}%
\bibitem [{\citenamefont {Haberman}\ and\ \citenamefont
  {Morton}(1953)}]{haberman1953}%
  \BibitemOpen
  \bibfield  {author} {\bibinfo {author} {\bibfnamefont {W.~L.}\ \bibnamefont
  {Haberman}}\ and\ \bibinfo {author} {\bibfnamefont {R.~K.}\ \bibnamefont
  {Morton}},\ }\href@noop {} {\emph {\bibinfo {title} {An experimental
  investigation of the drag and shape of air bubbles rising in various
  liquids}}},\ \bibinfo {type} {Tech. Rep.}\ (\bibinfo  {institution} {David
  Taylor Model Basin Rep. no 802},\ \bibinfo {year} {1953})\BibitemShut
  {NoStop}%
\bibitem [{\citenamefont {Saffman}(1956)}]{saffman1956}%
  \BibitemOpen
  \bibfield  {author} {\bibinfo {author} {\bibfnamefont {P.~G.}\ \bibnamefont
  {Saffman}},\ }\bibfield  {title} {\bibinfo {title} {On the rise of small air
  bubbles in water},\ }\href@noop {} {\bibfield  {journal} {\bibinfo  {journal}
  {J. Fluid Mech.}\ }\textbf {\bibinfo {volume} {1}},\ \bibinfo {pages} {249}
  (\bibinfo {year} {1956})}\BibitemShut {NoStop}%
\bibitem [{\citenamefont {Hartunian}\ and\ \citenamefont
  {Sears}(1957)}]{hartunian1957}%
  \BibitemOpen
  \bibfield  {author} {\bibinfo {author} {\bibfnamefont {R.~A.}\ \bibnamefont
  {Hartunian}}\ and\ \bibinfo {author} {\bibfnamefont {W.~R.}\ \bibnamefont
  {Sears}},\ }\bibfield  {title} {\bibinfo {title} {On the instability of small
  gas bubbles moving uniformly in various liquids},\ }\href@noop {} {\bibfield
  {journal} {\bibinfo  {journal} {J. Fluid Mech.}\ }\textbf {\bibinfo {volume}
  {3}},\ \bibinfo {pages} {27} (\bibinfo {year} {1957})}\BibitemShut {NoStop}%
\end{thebibliography}%

\end{document}